\documentclass{desyproc}

\begin{document}
\title{Unconventional Ideas for Axion and Dark Matter Experiments}

\author{{\slshape  Fritz Caspers}\\[1ex]
CERN, Geneva, Switzerland}

\contribID{fritz_caspers}

\confID{11832}  
\desyproc{DESY-PROC-2015-02}
\acronym{Patras 2015} 
\doi  

\maketitle

\begin{abstract}
In this contribution an entirely different way compared to conventional approaches for axion, hidden photon and dark matter (DM) detection is proposed for discussion. The idea is to use living plants which are known to be very sensitive to all kind of environmental parameters, as detectors.  A possible observable in such living plants could be the natural bio-photon level, a kind of metabolism related chemoluminescence. Another observable might be morphological changes or systematic leave movements. However a big problem for such kind of experiment would be the availability of a known, controllable and calibrated DM source. The objective of this small paper is primarily to trigger a debate and not so much to present a well-defined and clearly structured proposal.
\end{abstract}

\section{Introduction}
There appears to be growing evidence that very faint photon emissions in living biologic systems (bio-photons) could play an important role in intracellular communication and control of the growth. Those photon emissions are powered by the metabolism and can be considered as a kind of bio-luminescence. Dead plants or other dead organic materials do \emph{not} emit bio-photons. This kind of ``living cell radiation'' has been first postulated by A.~Gurwitsch nearly 100 years ago \cite{gurwitsch} and he conducted probably the first near UV  light shining through the wall experiment on onion roots. Of course everybody was laughing at him at this time. Around 1970 this kind of very faint radiation (range from 200 to 800~nm) on living plants was measured for the first time by F.A.~Popp \cite{popp} in Marburg (Germany) with highly sensitive photodetectors. Popp proposed that this radiation might be both semi-periodic and coherent. However this view is not generally accepted. If confirmed true, those biological systems (plants, cell cultures, etc.)  can react on very faint photon signals in a measurable way. We have two possible observables: observation of structure changes under the influence of some DM or axion flux (do we know it and are we able to control it?) and~/~or observation of changes of the bio-photon activity. With modern highly sensitive photon detectors and cameras the observation of those bio-photon activity is real fun and rather easy and one can see very nicely when e.g., some leaf of a plant is killed by injecting some toxic substance, how the bio-photon activity first strongly increases and then a few seconds to minutes later stops entirely (cry before death). But why should we consider to use plants or cell cultures and not observe such axion and DM related photons directly? Plants and cell cultures are full of cellular membranes (dielectric double layers and cell membranes with strong internal electric fields) where axions and other DM might convert into mm wave or probably optical photons which then could have an impact on the biological activity. Living systems are not in thermo-dynamical equilibrium (otherwise they would be dead) and thus may have a rather low effective ``noise temperature" (some people claim effects like stochastic resonance there). It should be noted that an electronic amplifier may exhibit a noise temperature of say 30~K when operating at ambient (a typical satellite antenna pre-amplifier for 10~GHz). The electronic amplifier is also not in thermo-dynamical equilibrium since it is connected to a power source. Perhaps such kind of bio-detectors are much more broadband that our presently used or discussed structures and they can be operated also in a strong magnetic field but of course not at cryo. There exist interesting theories by Fr\"ohlich \cite{froehlichMM,mayburovComm} on biological very low level coherent mm waves in biological systems.

\section{Designing an experiment}
Now regarding a practical proposal for such kind of test, I would propose to copy-paste one of the many plants experiments on temporal variation of the tidal force. The results are very convincing and also well accepted by the biologist's community \cite{barlowMoon,normann}. Observables are, amongst other parameters, leaf movements and morphological changes in the roots. Unfortunately we cannot just look for other, probably DM related periodicities in the observables since the known or anticipated variation of the DM flux (diurnal period) are extremely small. Lacking a controllable calibrated axion/hidden photon DM source we should maybe consider placing such an experiment in the vicinity of a nuclear reactor or beyond the (ionizing radiation) shielding of a beam dump/target which is frequently used, in some accelerator. It is clear that this experiment would NOT try to compete with biological experiments which are looking for ionizing radiation related changes e.g., on DNS strings or cell cultures. Such experiments have been carried out also in underground areas, but with negative results. Here the idea is rather to focus on observable ``behavioural'' or ``state'' changes e.g., leave movements and variations of the biophoton level. In any case the DM will not be seen (if any) directly by the plant but only via some real photons (mm wave range?) created by some conversion mechanism when e.g., hidden photons pass through cell membranes.

\section*{Acknowledgements}
The author would like to thank K.~Zioutas, M.~Schumann, E.~Wagner, P.~W.~Barlow and L.~Beloussov for stimulating and challenging discussions as well as M.~Betz for help in editing the manuscript.


\begin{footnotesize}

\end{footnotesize}



\begin{thebibliography}{99}

\bibitem{gurwitsch}
	A.~Gurwitsch,
	``Die Mitogenetische Strahlung. Monographien aus dem Gesamtgebiet der Physiologie der Pflanzen und der Tiere'',
	J. Springer, Bd. 25, Berlin (1932).
	
\bibitem{popp}
	F.~A.~Popp,
	``Properties of biophotons and their theoretical implications'',
	Indian Journal of Experimental Biology, \textbf{41}, 391, (2003).

\bibitem{froehlichMM}
	H.~Fr\"ohlich,
	``Coherent Electrical Vibrations in Biological Systems and the Cancer Problem'',
	IEEE Transactions on Microwave Theory and Techniques, VOL. MIT-26, NO. 8, (1978).

\bibitem{mayburovComm}
	S.~N.~Mayburov,
	``Photonic Communications and Information Encoding in Biological Systems'',
	arXiv:1205.4134,
	Quantum Information conference, Torino, (2012).
	
	
\bibitem{barlowMoon}
	P.~W.~Barlow \textit{et al.},
	``Arabidopsis thaliana root elongation growth is sensitive to lunisolar tidal acceleration and may also be weakly correlated with geomagnetic variations'',
	Annals of Botany \textbf{111}, 859, (2013).
	
\bibitem{normann}
	J.~Normann \textit{et al.},
	``Rhythms in Plants'',
	Springer, ISBN 978-3-319-20516-8, pp. 35-55, (2015).
	
	

\end{thebibliography}
\end{document}